\documentclass[prl,showpacs,aps]{revtex4}
\usepackage{graphicx}
\usepackage{epsfig}

\begin{document}

\title{\rm Optimization of Partial Search}
\author{ Vladimir\ E.\ Korepin}
\affiliation{
C.N. Yang Institute for Theoretical Physics, State
 University of New York at Stony Brook,\\
 Stony Brook, NY 11794-3840
e-mail: korepin@insti.physics.sunysb.edu}
\date{\today}

\bigskip

\begin{abstract}

Quantum Grover search  algorithm can find a target item in a database faster
than any classical algorithm. One can trade accuracy for speed and find a
part of the database (a block) containing the target item even  faster, this 
is  partial search.  A partial search algorithm was recently suggested by
Grover and Radhakrishnan. Here we  optimize it.
Efficiency  of the search algorithm is measured by number of 
queries to the oracle. The author suggests new version of
Grover-Radhakrishnan  algorithm which uses minimal number of queries to
the oracle. The algorithm can run  on the same hardware which is used for the
usual Grover algorithm.
\end{abstract}


\maketitle

\section{Introduction}
Database search has many applications and  used widely.  
Grover discovered a quantum algorithm that searches faster than a classical
algorithm \cite{Grover}.  It consists of repetition of the Grover iteration
$G_1$. We shall call it global iteration, see (\ref{iter}).
The number of  repetitions is:
\begin{equation}
j_{\mbox{full}}= \frac{\pi}{4} \sqrt{N} \label{full}
\end{equation}
for a database with large number of entries $N$. After $j_{\mbox{full}} $
the algorithm finds the target item. 

Sometimes  it is sufficient to find an approximate location of  the target
 item. A partial search considers the following problem: a  database is
 separated into   
$K$ { blocks},
of  a  size
$b={N}/{K}$.
We  want to find a block with the target item, not the target  item itself. 
First  quantum algorithm for a partial search was suggested  by Grover and 
 Radhakrishnan in \cite{jaik}.
They showed that  classical partial search takes  $\sim (N-b)$ queries, but
quantum  algorithm takes only 
$\sim ( \sqrt{N}- {\bf c}\sqrt{b}) $ queries. Here $\bf c$
is a positive coefficient.
This algorithm  uses several global iterations $G^{j_1}_1$ and then several 
local  iteration $G^{j_2}_2$, see (\ref{liter}). Local searches are made in 
each  block separately in parallel.
Here we optimize this algorithm:
the number of  queries to the oracle  minimized, the coefficient  $\bf c$ is 
 increased.
Exact expression  for the number of queries necessary to find the target block is given by formulae  (\ref{answer}),(\ref{shag}) and  (\ref{zapros}).
Efficiency of search algorithms is measured by number of 
queries to the oracle, we call it number of iterations. The lower bound is in 
the end of the  paper.
Partial search can use the same hardware as the full search.

\section{Partial Search}
\subsection{Global Iterations}
First let us remind the full Grover search.
We  consider a database with one target item.
The aim of the Grover algorithm is to identify a target state $|t\rangle$
among an unordered set of $N$ states.
This is achieved by repeating global iteration which is defined in terms of 
two operators. The first changes the sign of the target state $|t\rangle$ only:
\begin{equation}
I_t=\hat{I}-2|t\rangle \langle t|, \qquad \langle t|t\rangle=1, 
\label{target}
\end{equation}
where $\hat{I}$ is the identity operator. 
The second operator, 
\begin{equation}
I_{s_1}=\hat{I}-2|s_1\rangle \langle s_1|, \label{average}
\end{equation}
changes the sign of the uniform superposition of all basis states
$|s_1\rangle$, 
\begin{equation}
|s_1\rangle = \frac{1}{\sqrt{N}}\sum_{x=0}^{N-1}|x\rangle , \qquad 
\langle s_1|s_1\rangle =1 . \label{ave}
\end{equation}
The {\bf global iteration} is defined as a unitary operator
\begin{equation}
G_1=-I_{s_1}I_t . \label{iter}
\end{equation}
We shall use  eigenvectors  of $G_1$: 
\begin{eqnarray}
 G_1|\psi^{\pm}_1\rangle  = \lambda^{\pm}_1 |\psi^{\pm}_1\rangle , \qquad 
\lambda^{\pm}_1 =\exp[{\pm 2i \theta_1}] , \qquad
 |\psi^{\pm}_1\rangle  = \frac{1}{\sqrt{2}}|t\rangle \pm \frac{i}
{\sqrt{2}}\left( \sum^{N-1}_{\stackrel{\mbox{\small{x=0}}}{\mbox{\small{x  $\neq$  t}}}} 
\frac{|x\rangle}{\sqrt{(N-1)}} \right). \label{value}
\end{eqnarray} 
They were found in \cite{Brass}.
The angle $\theta_1$ is defined by 
\begin{equation}
\sin ^2\theta_1 =\frac{1}{N}. \label{ang1}
\end{equation}


\subsection{Grover-Radhakrishnan Algorithm for Partial Search}

The partial search algorithm is designed to find a block with the target 
item: the target block. We shall call other blocks: non-target blocks.
The algorithm uses $j_1$ global iteration and $j_2$ local iterations.
Local iterations  are Grover iterations for each block:
\begin{equation}
G_2=-I_{s_2}I_t .\label{liter}
\end{equation}
$I_t $ is given by (\ref{target}), but $I_{s_2}$ is different.  The action of
the operator   $I_{s_2}$  in an individual  block can be
 represented as:
\begin{equation}
I_{s_2}\mid_{\mbox{\scriptsize{block}}}=\hat{I}\mid_{\mbox{\scriptsize{block}}}-2|s_2\rangle \langle s_2|, \qquad
|s_2\rangle = \frac{1}{\sqrt{b}}\sum_{\mbox{\scriptsize{the block}}}|x\rangle .
\label{nblock}\end{equation}
For the whole database we should write  $I_{s_2}$
as a direct sum of the operators (\ref{nblock}) 
over all blocks.

Relevant eigenvectors of $G_2$ are:
\begin{eqnarray}
 G_2|\psi^{\pm}_2\rangle =\lambda^{\pm}_2 |\psi^{\pm}_2\rangle, \qquad
 \lambda^{\pm}_2  = \exp[{\pm 2i \theta_2}] , \label{vector2}  \qquad
|\psi^{\pm}_2\rangle   =\frac{1}{\sqrt{2}}|t\rangle \pm \frac{i}{\sqrt{2}}
|\mbox{ntt}\rangle
\end{eqnarray} 

Here the $|\mbox{ntt}\rangle $ is a normalized sum of all non-target items 
in the target block:
\begin{equation}
|\mbox{ntt}\rangle =
\frac{1}{\sqrt{b-1}} \sum_{\stackrel{x \neq t}{\mbox{\tiny{target block}}}}
|x\rangle,\qquad \langle \mbox{ntt} |\mbox{ntt}\rangle =1 .\label{vntt}
\end{equation}
We shall need  an angle $\theta_2$ given by
\begin{equation}
\sin^2 \theta_2 =\frac{K}{N}=\frac{1}{b}. \label{ang2}
\end{equation}
Local iteration does not change non-target blocks.
Inside the target block it acts similar to the usual Grover search.
After several global iterations and several local we still have to apply 
one more global iteration. The partial search  algorithm  creates a vector
\begin{equation}
|d\rangle =G_1 G^{j_2}_2 G^{j_1}_1|s_1\rangle .\label{gl}
\end{equation}
In the state $|d\rangle $ the amplitudes of all items in non-target blocks 
are the same. Using eigenvectors of local (\ref{vector2}) and global iterations from (\ref{value}) we 
can calculate this amplitude and require that it vanishes:
\begin{eqnarray}
&\frac{-N}{\sqrt{N-1}} \left( \frac{1}{2} -\frac{1}{K} \right)\cos \left( (2j_1+1)\theta_1 \right) = 
\cos (2j_2\theta_2) \sin \left( (2j_1+1)\theta_1 \right) + \sqrt{\frac{b-1}{N-1}} 
\sin (2j_2\theta_2)\cos \left( (2j_1+1)\theta_1 \right)  \nonumber  \\
& -\sqrt{b-1}\sin (2j_2\theta_2) \sin \left( (2j_1+1)\theta_1 \right) + {\frac{b-1}{\sqrt{N-1}}} \cos (2j_2\theta_2)\cos \left( (2j_1+1)\theta_1 \right) \label{center}
\end{eqnarray}
This equation  guarantees that  the  amplitude of each item in each 
non-target block vanishes.
Now we can  {\it measure}. In the simplest case $N=2^n$ and $K=2^k$, so  we can 
 label blocks by $k$ qubits [items inside of a block are labeled by $n-k$ qubits]. We measure only $k$ block  qubits and 
{\bf  find the  target block}.
We shall choose the numbers of iterations $j_1$ and $j_2$ by minimizing 
the total number of iterations  $j_1+j_2 $.

To see universal features  we consider the limit when  each block is very large
$b\rightarrow \infty$, this makes the total number of items in the whole database   also large
 $ N=Kb\rightarrow \infty$.
 The expression for  
angles (\ref{ang1}), (\ref{ang2}) simplifies: 

$\theta_1={1}/{\sqrt{N}},\quad \theta_2={1}/{\sqrt{b}}.$
It was shown in \cite{jaik} that the numbers of iterations  scales   as:
\begin{equation}
j_1=\frac{\pi}{4}\sqrt{N} -\eta \sqrt{b} ,\qquad j_2=\alpha 
\sqrt{b}, \qquad {\bf c}=\eta -\alpha. \label{steps}
\end{equation}
Here $\eta$ and $\alpha$ are parameters of order of 1 [they have a  limit]. 
For large blocks $b\rightarrow \infty$ the equation (\ref{center})  can be 
simplified to:
\begin{equation}
\tan \left(\frac{2\eta}{\sqrt{K}} \right) =\frac{2\sqrt{K}\sin 2\alpha}{K
 -4\sin^2 \alpha }  \label{expr}
\end{equation}

\subsection{Minimization of Total Number of Iterations. }
Let us minimize the  number of queries to the oracle [number of iterations]:
$S=j_1+j_2+1\rightarrow \frac{\pi}{4}\sqrt{N}  -{\bf c}  \sqrt{b}.$
Here ${\bf c}=\eta- \alpha $
To optimize the algorithm we have to minimize 
$(\alpha -\eta) $ having in mind constrain (\ref{expr}).
The author found the optimal values of $\alpha $ and $\eta$, they 
depend on $K$, let us distinguish them
by a subindex $\alpha_K $ and $\eta_K$.
The minimum number of queries   is achieved at:
\begin{equation}
\tan \frac{2\eta_K}{\sqrt{K}}=\frac{\sqrt{3K-4}}{K-2}, \qquad    
\cos 2\alpha_K=\frac{K-2}{2(K-1)}, \qquad   {\bf c}=\eta_K - \alpha_K  . \label{answer}
\end{equation}
 {\bf This describes  optimal version of Grover-Radhakrishnan algorithm}.

Let us study dependence on number of blocks $K$:
 $\alpha_K $ monotonically decreases
with  $K$:
\begin{eqnarray}
\alpha_2=&\frac{\pi}{4}\ge \alpha_K &\ge  \frac{\pi}{6} =\alpha_{\infty} \label{range}\\
 K=2 & \rightarrow & K=\infty \nonumber
\end{eqnarray}
 In case of  two large blocks $K=2$
minimization of number of queries of partial search algorithm  gives:
$\alpha_2={\pi}/{4}$ ,  $\eta_2= {\pi}/{2\sqrt{2}}. $
This means that for $K=2$ algorithm skips the global iterations and  makes a
 full local  search in  each block: 
$ j_1=0,  \quad   j_2 = ({\pi}/{4}) \sqrt{b}. $
For three blocks  or more   $3\le K$ the  algorithm makes  less then full  
 search of each block [locally].
Now let us analyze the number of global iterations:
\begin{equation}
j_1= \left(  \frac{\pi}{4}- \frac{\eta_K}{\sqrt{K}}  \right)\sqrt{N}>0, \quad
\frac{d}{dK}\left(\frac{\eta_K}{\sqrt{K}}\right)<0, \quad  \frac{dj_1}{dK}>0, \quad \mbox{for} \quad  3\le K .
\end{equation}
Parameter $\eta_K$  decreases monotonically   from
$\eta_2= {\pi}/(2\sqrt{2})$ to $\eta_{\infty}=\sqrt{3/4}$, when $K$ increases.

The difference $\alpha_K-\eta_K$ monotonically decrease with $K$.
Numerical values  of  $\alpha_K$ and  and $\eta_K $ for different number of 
blocks  are:
\begin{eqnarray}
&\alpha_2 \approx  0.7854 ,&\quad \eta_2 \approx  1.1107 ,  \quad \alpha_2 -\eta_2 \approx -0.3253
 \nonumber  \\
&\alpha_3 \approx 0.65906,&\quad \eta_3 \approx 0.9961 ,  \quad \alpha_3 -\eta_3 \approx -0.33704\nonumber  \\
&\alpha_4 \approx  0.6155,& \quad  \eta_4 \approx  0.9553 , \quad 
\alpha_4 -\eta_4 \approx -0.3398 \nonumber  \\
&\alpha_5 \approx 0.5932,&\quad  \eta_5 \approx 0.9341, \quad \alpha_5 -\eta_5 \approx -0.3409 \nonumber \\
&\alpha_{\infty}\approx 0.5236,&  \quad \eta_{\infty}\approx 0.866,\quad  \alpha_{\infty} -\eta_{\infty}\approx -0.3424 \nonumber
\end{eqnarray}
These are solutions of equation (\ref{answer}).
{These  parameters   define the number of iterations 
\begin{eqnarray}
&  j_1=\frac{\pi}{4}\sqrt{N} -\eta_K \sqrt{b} ,\qquad j_2=\alpha_K 
\sqrt{b},  \quad S_K \approx j_1+j_2\rightarrow \frac{\pi}{4}\sqrt{N} +(\alpha_K -\eta_K) 
 \sqrt{b} . \label{shag} 
\end{eqnarray}
We can compare this with the full search in {\it randomly} picked $K-1$ blocks, which
takes
\begin{equation}
R_K=\frac{\pi}{4}\sqrt\frac{K-1}{K}\sqrt{N} \label{random}
\end{equation}
iterations,  see (\ref{full}). 
For two blocks partial search and random pick takes the same number of queries:
$R_2= S_2= [{\pi}/{4}]\sqrt{{N}/{2}}.$
For more blocks  partial search is faster:
\begin{eqnarray}
&R_3=0.641\sqrt{N}, \qquad  &S_3=0.59\sqrt{N}, \nonumber \\
&R_4=0.68\sqrt{N}, \qquad  &S_4=0.586\sqrt{N}, \nonumber \\
&R_5=0.702\sqrt{N},\qquad   &S_5=0.63\sqrt{N}. \nonumber 
\end{eqnarray}
Here we compared random pick algorithm with the partial search algorithm using:
$S_K= \left( {\pi}/{4} +[{{\alpha_K -\eta_K}]/{\sqrt{K}}}\right) 
\sqrt{N}. $
We see that starting from $K=3$ partial search algorithm works faster then random pick.
As the number of blocks increases the advantage becomes more essential.

But for large $K$ we should  compare the partial search algorithm with its {\it interrupted} version:
If we  make only global iterations  of the partial search algorithm and
 measure the wave function of the  database, 
probability to find the target item is:
\begin{equation}
p_t=\sin^2 \left( (2j_1+1)\theta_1 \right)=\frac{(K-2)^2}{K(K-1)}. \label{abort}
\end{equation}
It monotonically increases with $K$.

 Let us solve equations 
(\ref{answer})  explicitly for large $K$:
\begin{eqnarray}
\alpha_{K} &\rightarrow \frac{\pi}{6}+\frac{1}{2\sqrt{3} K}+
 \frac{5\sqrt{3}}{(6K)^2}, \label{depend} \qquad  \eta_{K} &\rightarrow  \frac{\sqrt{3}}{2}+\frac{1}{2\sqrt{3} K} +
\frac{11\sqrt{3}}{90K^2} . 
\qquad   K\rightarrow \infty   \nonumber 
\end{eqnarray}
Corrections to these expressions are of order $1/K^3$.
The total number of queries is:
\begin{eqnarray}
S_K \rightarrow \frac{\pi}{4}\sqrt{N} +(\alpha_K -\eta_K) 
 \sqrt{b} , \label{total} \qquad -{\bf c}=\alpha_K -\eta_K=  \frac{\pi}{6}-
\sqrt{\frac{3}{4}} +\frac{1}{5\sqrt{3}(2K)^2} <0. 
\qquad 
  \label{zapros}
\end{eqnarray}

 Random pick  (\ref{random}) takes more queries:
\begin{equation}
R_K\rightarrow \frac{\pi}{4}\sqrt{N} -\left( \frac{\pi}{8\sqrt{K}}
 \right) \sqrt{{b}}, \qquad K \rightarrow \infty .
\end{equation}
As for the {\it interrupted} version of the algorithm in the limit of $K\rightarrow \infty$,
the probability to  find the target item by measuring after global iterations
 is close  to certainty:
$p_t= 1-{3}/{K},\quad K\rightarrow \infty $
see (\ref{abort}). The partial search algorithm is efficient for limited
 number of blocks only:
$3\le K \le {3}/(1-p_t).$
 If we  choose the probability  $p_t=0.9$ then the partial
 search algorithm works well in the region:
\begin{equation}
3\le K \le 30. \label{trid}
\end{equation}

The version of  partial search algorithm described here  is   little faster 
then original Grover-Radhakrishnan algorithm \cite{jaik}: in the expression for
total number of iterations $S_K$ the coefficient  ${\bf c}=\eta_K-\alpha_K$
in (\ref{zapros}) and (\ref{shag}) is
 from 1$\%$ to 3$\%$  larger [depending on $K$]. But our  version  uses
 the absolute minimum  of queries to the  oracle.

\subsection{Lower bound}
A lower bound for number of queries to the oracle was found in \cite{jaik}:
\begin{equation}
S\ge \frac{\pi}{4}\sqrt{N}-\frac{\pi}{4}\sqrt{b}.
\end{equation}
It is based on the lower bound for the full search \cite{Bennett,zalka}.
One can first search for the block and then for the target item in the block.
We can improve the lower bound for algorithms that have the same final state
for the target block. After we run partial search  algorithm the wave 
function of the database (\ref{gl}) has non-zero components only in the target block. The calculations show:
\begin{eqnarray}
 |d\rangle= \sin  \alpha_K|t\rangle &+&\cos  \alpha_K |\mbox{ntt} \rangle \label{eff} 
\end{eqnarray}
see (\ref{answer}) and (\ref{vntt}).  We can
represent it as a result of application of $j_e$ Grover iterations to 
uniform superposition of all basis states in the target block:
\begin{eqnarray}
 |d\rangle= G_2^{j_{e}}|s_2 \rangle ,\qquad j_e= \frac{\alpha_K}{2}\sqrt{b} \label{usk} 
\end{eqnarray}
see (\ref{liter}) and (\ref{nblock}).
It will take only 
$\tilde{j}_{\mbox{full}}=\left({\pi}/{4}-{\alpha_K}/{2}\right)\sqrt{b}$
iterations
to find the target item  in the target block.
We can a  bound  $S$ from the following:
$S+\tilde{j}_{\mbox{full}} \ge {\pi}\sqrt{N}/4$.
Lower bound depends on number of blocks, see (\ref{range}).
Replacing $\alpha_K$ by its minimum (\ref{range}) we get a tighter lower bound:
\begin{equation}
S\ge \frac{\pi}{4}\sqrt{N}+ \left(-\frac{\pi}{4}+\frac{\alpha_K}{2}\right)\sqrt{b} \ge \frac{\pi}{4}\sqrt{N}-\frac{\pi}{6}\sqrt{b}.
\end{equation}

\section{Summary}
We optimized Grover-Radhakrishnan version of partial search.
We conjecture that our version of partial search is optimal in wider class of
partial search algorithms [arbitrary sequences of local and global iterations].

\section{Acknowledgments}
The author is  grateful  to   L.K Grover and J. Radhakrishnan  for
 productive  discussions.


\end{document}